\documentclass[onecolumn,secnumarabic,amssymb, nobibnotes, aps, prd]{revtex4-2}

\usepackage{comment}
\usepackage{graphicx}
\usepackage{amsmath}
\usepackage{mathtools}
\usepackage{slashed}
\usepackage{xcolor}
\usepackage[force]{feynmp-auto}
\usepackage{grffile}
\usepackage{float}
\usepackage{bbm}

\begin{document}

\title{Dark sector tensor currents contribution to lepton's anomalous magnetic moment}%

\author{V. Kozhuharov, M. Naydenov}%
\email{mnaydenov@phys.uni-sofia.bg}
\affiliation{Faculty of Physics, Sofia University, 5 J. Bourchier Blvd, 1164 Sofia, Bulgaria}
\date{}%
%\tableofcontents

\begin{abstract}

In this work we consider a model including dark sector bosons interacting through tensor currents with Standard Model leptons. We show that for certain values of the interaction constant this model has the potential of providing an explanation for the discrepancy between theory and experiment, regarding the anomalous magnetic moment of the muon.
%as well as the $^7$Be and $^4$He decays. 
The effect on the already established measurements for the electron 
are small and the lepton universality violation is naturally incorporated. Possible experimental searches together with systematic approach to 
characteristic properties of the final states are discussed.
%for spectral differentiation between the different end products.
  
\end{abstract}

\maketitle

\section{Overview}

Quantum electrodynamics is the theory giving the most accurate predictions confirmed by experiment in the history of physics. One of the experimentally testable consequences of the theory is the prediction first made by Julian Schwinger for the anomalous magnetic moment of the electron \cite{Schwinger}. The anomalous magnetic moment is a quantum property of charged leptons arising from loop corrections to the fermionic electromagnetic vertex. It is defined as $a=\frac{g-2}{2}$, where $g$ is the Land\'e g-factor and the 
agreement between theoretical predictions and experiment for the case of an electron, including corrections from QED, hadron physics and electroweak theory is \cite{AMME}

\begin{equation}
\Delta a_e=a_e^{\text{EXP}}-a_e^{\text{SM}}=(4.8 \pm 3)\times10^{-13}.
\end{equation}

%\noindent 
A problem arises when one tries to treat the muon or the tau lepton the same way as the electron. The discrepancy between theory and experiment for the case of the muon is \cite{AMMM}

\begin{equation}
\Delta a_\mu=a_\mu^{\text{EXP}}-a_\mu^{\text{SM}}=(251\pm59)\times10^{-11}.
\end{equation}

%\noindent 
One sees that the application of the same theoretical treatment to leptons from different families leads to different accuracy of the prediction and the lepton universality violation is manifest. There exist independent experimental evidences supporting this difference  \cite{ex1, ex2}. 
This is one of the main motivations for searching for physics beyond the Standard Model which potentially compensates for the discrepancy between theory and experiment. 

Various probable solutions to the $\Delta a_\mu$ problem have been suggested - possibly within the Standard Model \cite{SM1}, including leptoquarks \cite{leptoquarks}, using supersymmetric models \cite{supersymmetry}, and various other exotic models \cite{nonlocalQED, gravitino}. Others consider high energy solutions, testable at the LHC \cite{model1, model2}. So far, no generally applicable solution to the muon anomalous magnetic moment problem
exists that is also supported by the experimental observations.
In addition to the $g-2$ problem 
%we would also like to address the 
 the
$^8Be$ decay anomaly observed at the ATOMKI collaboration \cite{atomki}
%which 
might as well be a strong sign for physics beyond the Standard Model \cite{rev1, rev2, rev3}.

%which does not disturb on its own results already confirmed by experiment.

In this work we are particularly interested in solutions involving dark sector particles, similar to the models presented in \cite{darkscalar, darkvector}. 
The dark sector extension was proposed by B. Holdom \cite{holdom} where the U(1) gauge group is generalized to U$'$(1) by adding a massive dark gauge particle interacting with visible matter through an interaction constant related to the charge of the electron. 
Then the inclusion of dark scalar, pseudo-scalar and pseudo-vector terms in the Lagrangian is trivial. 
Only the pseudo-scalar and the vector particles can serve as a solution to the ATOMKI anomaly. 

%In a previous work we showed \cite{MomchilVenelin} that dark boson states are mathematically admissible and are eventual candidates for explaining the anomaly observed by the ATOMKI collaboration.

%Since these dark sector states can potentially describe the unexpected beryllium decay, an obvious continuation of our work was to check if the predicted constant of interaction between ordinary matter and dark tensor bosons is compatible with the discrepancy between theory and experiment for the muon anomalous magnetic moment. Further condition is that the influence on the electron anomalous magnetic moment must be small. 

In the present paper a model Lagrangian will be discussed and the corresponding 
corrections to the photon vertex will be numerically estimated. 
We show that for the lightest charged lepton (electron) the correction to the anomalous magnetic moment can be negligibly small compared to the established experimental accuracy, and the heavier the lepton is, the bigger is the correction to the anomalous magnetic moment. 
%In this sense, the introduction of tensor currents has the potential of explaining two contemporary theoretical problems at once - the atypical beryllium decay and the mismatch between theory and experiment for the anomalous magnetic moment of the muon.

This paper is organized as follows: we start with a Lagrangian describing the phenomenology of fermion states interacting through tensor currents, we derive the consequent Feynman rules for the various vertices and the resulting propagators and then we calculate the resulting correction 
%that such particles give 
to the electromagnetic vertex. We show that dark tensor bosons can 
be responsible for 
%both 
the muon $g-2$ anomaly, and have the potential to 
be discovered at present and future experiments
%the beryllium decay. Accelerator experiments, 
such as PADME \cite{bib:padme}, SeDS \cite{bib:seds}.
%, and ALICE \cite{bib:alice}.
%can be sensitive to effects predicted by the proposed model.

\section{The model}

The Lagrangian proposed in the paper by B. Holdom is an extension of the U(1) including a dark massive vector particle, the dark photon. This Lagrangian can be extended for fields described by Lorentz invariant currents corresponding to scalar, pseudo scalar, pseudo vector, tensor and pseudo tensor terms. We are especially interested in processes which have the potential of explaining the $^8$Be anomaly
%which limits the spectrum of available solutions to massive vector, tensor and pseudo tensor fields. 
by considering the Lagrangian

\begin{equation}
\begin{gathered}
    \mathcal{L}=-e\overline{\Psi}\gamma_\mu A^\mu\Psi-e_1\overline{\Psi}\gamma_\mu A_1^\mu\Psi-e_2\overline{\Psi}\gamma_\mu\gamma^5A_2^\mu\Psi+ie_3\overline{\Psi}\gamma^5A_3\Psi-ie_4\overline{\Psi}\frac{q^\mu}{|q|}\sigma_{\mu\nu}A_4^\nu\Psi+e_5\overline{\Psi}\frac{q^\mu}{|q|}\sigma_{\mu\nu}\gamma^5A_5^\mu\Psi+\\
    +i\overline{\Psi}\gamma^\mu\partial_\mu\Psi-m\overline{\Psi}\Psi.
    \end{gathered}
\label{eq:gen-lagr}
\end{equation}

%\noindent 
The first term is the lepton-photon interaction from standard QED and then we have in turn the interaction with dark sector particles - a vector, pseudo vector, pseudo scalar, tensor and pseudo tensor current. 
At the end we have the kinetic terms for the lepton of interest. The factors $e_i$ are dimensionless constants of interaction. Gauge fields can be taken into account by including 1-loop corrections with four external lines. The chiral symmetry condition requires the definition of the tensor currents to include the incoming momentum of the particle $q$. 

A tensor vertex  was proposed by 
Nambu and Jona-Lasinio \cite{NJL} for the interactions between mesons and fermions,  
in analogy with superconductivity. 
The trivial extension of vector currents to tensor currents leads to a chiral pair of terms $(\Bar{\Psi}\sigma_{\mu\nu}\Psi)^2+(\Bar{\Psi}i\gamma^5\sigma_{\mu\nu}\Psi)^2$, which for chiral transformations is identically 0. Therefore, the Lagrangian should contain a unique momentum dependence, because the local product of two tensor currents with different chiralities vanishes identically  \cite{chizhov,eguchi}. 
The appearance of a tensor vertex in eq. (\ref{eq:gen-lagr}) can be seen as an 
effective low-energy approximation (i.e. effective interaction) of a more 
fundamental theory, for example string field theories, where 
non-local interactions arise naturally due to 
the extended structure of the fundamental objects. 
%We exploit this idea for the case of dark meson, coupling to fermions and explore the consequences for the anomalous magnetic moment \cite{chizhov,eguchi}. 

After leaving only terms containing vector, tensor and pseudotensor interactions as in \cite{MomchilVenelin}
%having the potential of being experimentally detected 
we are left with the final Lagrangian which will be used 
throughout the paper: 

\begin{equation}
\begin{gathered}
    \mathcal{L}=-e\overline{\Psi}\gamma_\mu A^\mu\Psi-e_1\overline{\Psi}\gamma_\mu A_1^\mu\Psi-ie_4\overline{\Psi}\frac{q^\mu}{|q|}\sigma_{\mu\nu}A_4^\nu\Psi+e_5\overline{\Psi}\frac{q^\mu}{|q|}\sigma_{\mu\nu}\gamma^5A_5^\nu\Psi+i\overline{\Psi}\gamma^\mu\partial_\mu\Psi-m\overline{\Psi}\Psi.
    \end{gathered}
\label{lagrangian}
\end{equation}

The Feynman rules emerging from this model are the following:

\begin{figure}[ht]
\includegraphics[width=0.8\textwidth]{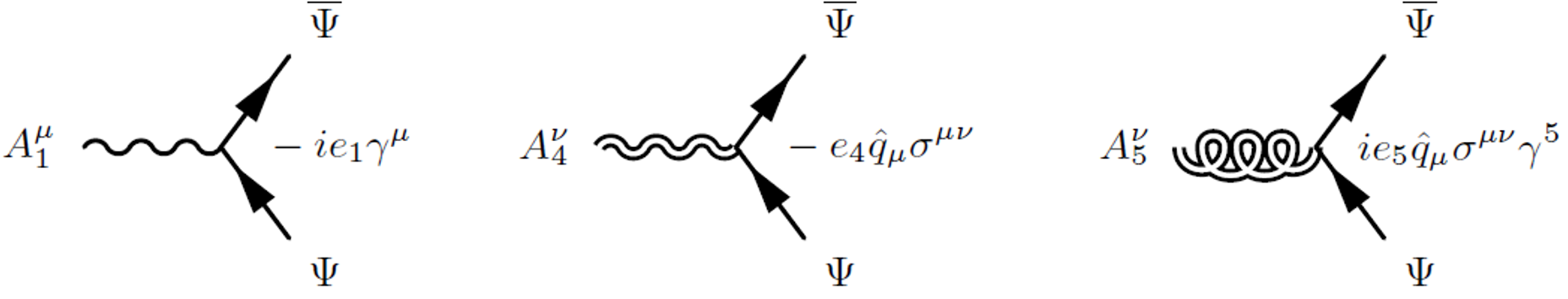}
\centering
\caption{Basic Feynman rules for leptonic interactions with dark sector particles. Here $\hat{q}^\mu=\frac{q^\mu}{|q|}$.}
\end{figure}

\begin{figure}[ht]
\includegraphics[width=0.8\textwidth]{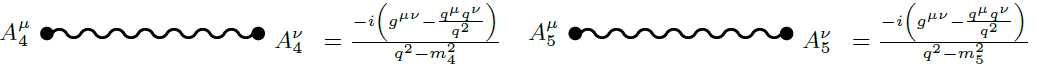}
\centering
\caption{Propagator rules for the $A_4^\mu$ and $A_5^\mu$ dark bosons.}
\label{fig:pi-ggp1}
\end{figure}

The magnitude of the constants of interaction are defined as a
rescaled electron charge $e_i=\epsilon_i e$, where $\epsilon_i$ 
is a rescaling factor governing the mixing between the photon and the dark sector particles.

\section{Anomalous magnetic moment contribution due to dark sector bosons}

%In a previous work \cite{MomchilVenelin} we show that the anomalous beryllium decay might be due to any of the dark particles included in this model and so far there are no experimental evidences leading unambiguously to one of them. 
%Therefore, following the results in \cite{MomchilVenelin} we assign the same interaction constant $e'=\epsilon e$. 
A possible existence of new vector particles interacting through 
(pseudo-)tensor currents with the fundamental leptons will 
modify their magnetic moment.
The subsequent calculations are performed for
independent values of $\epsilon_4$ and $\epsilon_5$ but 
where applicable, a typical benchmark value is used, $\epsilon_4=\epsilon_5=\epsilon = 10^{-3}$. % $\epsilon^2=3\cdot 10^{-9}.$ 
The diagrams influencing the anomalous magnetic moment of the leptons are the following:

\begin{figure}[ht]
\includegraphics[width=0.5\textwidth]{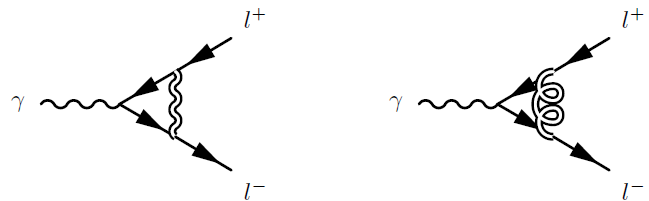}
\centering
\caption{Feynman diagrams for the electromagnetic correction due to $A_4^\mu$ and $A_5^\mu$ dark bosons.}
\label{fig:pi-ggp2}
\end{figure}
%\noindent 
The electromagnetic vertex function can be expanded in terms of form factors as
\begin{equation}
    \Gamma^\mu=F_1(k^2)\gamma^\mu+F_2(k^2)\frac{i\sigma^{\mu\nu}}{2m}k_\nu,
    \label{formfactor}
\end{equation}
where $k$ is the photon momentum. At tree level the electron is a point-like particle, where $F_1=1$ and $F_2=0$. Quantum corrections from 1-loop diagrams give rise to non-trivial behaviour of the form factors in which in standard QED $F_1(k^2)$ contains infrared divergence and the quantum contribution to the anomalous magnetic moment is evaluated when taking the limit $F_2(k^2=0).$

For the case of a virtual tensor boson $A_4^\mu$ we obtain
\begin{equation}
    \delta_4\Gamma^\mu(k)=-\epsilon_4^2e^2\int\frac{d^4q}{(2\pi)^4}\frac{1}{q^2}\frac{g_{\alpha\beta}}{q^2-m_4^2}\overline{u}(p_1)\sigma^{\alpha\rho}q_\rho\frac{\hat{p}_1-\hat{q}+m}{(p_1-q)^2-m^2}\gamma^\mu\frac{\hat{p}-\hat{q}+m}{(p-q)^2-m^2}\sigma^{\beta\omega}q_\omega u(p)
\end{equation}
\noindent and for the case of a virtual pseudo-tensor boson $A_5^\mu$

\begin{equation}
    \delta_5\Gamma^\mu(k)=-\epsilon_5^2e^2\int\frac{d^4q}{(2\pi)^4}\frac{1}{q^2}\frac{g_{\alpha\beta}}{q^2-m_5^2}\overline{u}(p_1)\sigma^{\alpha\rho}q_\rho\gamma^5\frac{\hat{p}_1-\hat{q}+m}{(p_1-q)^2-m^2}\gamma^\mu\frac{\hat{p}-\hat{q}+m}{(p-q)^2-m^2}\sigma^{\beta\omega}q_\omega\gamma^5 u(p).
\end{equation}

%\noindent 
These integrals are calculated using a \textit{Mathematica} package \cite{XPackage}, where the result is the contribution to the tensor part in Eq. (\ref{formfactor}). Here the dependence on $k$ is hidden in the kinematical relation between the momenta $p$, $p_1$ and $k$ and is made manifest by using Mandelstam variables and by the requirement for conservation of energy. Setting the condition $k^2=0$ we obtain the two corrections

\begin{equation}
\begin{gathered}
\delta_4\Gamma^\mu=\frac{e^2 \epsilon_4^2 }{16 \pi ^2}\Bigg(-\frac{9 m^2+2 M^2}{m^2}+\frac{\left(8 m^4-3 m^2 M^2-M^4\right) \ln \left(\frac{m^2}{M^2}\right)}{m^4}-\\
-\frac{2 \sqrt{M^2 \left(M^2-4 m^2\right)} \left(16 m^4-m^2 M^2-M^4\right) \ln \left(\frac{\sqrt{M^2-4 m^2}+M}{2 m}\right)}{m^4 \left(4 m^2-M^2\right)}\Bigg)
\end{gathered}
\end{equation}

\begin{equation}
\begin{gathered}
\delta_5\Gamma^\mu=\frac{e^2 \epsilon_5^2 }{16 \pi ^2}\Bigg(-\frac{3 m^2-2 M^2}{m^2}-\frac{M^2 \left(3 m^2-M^2\right) \ln \left(\frac{m^2}{M^2}\right)}{m^4}-\\
-\frac{2 \left(m^2-M^2\right) \sqrt{M^2 \left(M^2-4 m^2\right)} \ln \left(\frac{\sqrt{ \left(M^2-4 m^2\right)}+M}{2 m }\right)}{m^4}\Bigg),
\end{gathered}
\end{equation}

\noindent where we take $m_4=M$ or $m_5=M$ and $m$ is the lepton mass in the final state. 
%Plotted graphically, the accumulated effect on the anomalous magnetic moment $\delta_4\Gamma^\mu+\delta_5\Gamma^\mu$ is  

%\begin{figure}[ht]
%\includegraphics[width=0.5\textwidth]{correction_plot.jpg}
%\centering
%\caption{The change correction to the lepton anomalous magnetic moment due to dark sector tensor bosons.}
%\end{figure}

%\noindent 
We obtain a total correction to the anomalous magnetic moment as a function of the lepton mass. One can note that the correction increases for bigger masses, so the influence on the electron magnetic moment is negligible, and for the muon and potentially the tau lepton is much bigger. %, as desired. 

\begin{figure}[ht]
\includegraphics[width=0.45\textwidth]{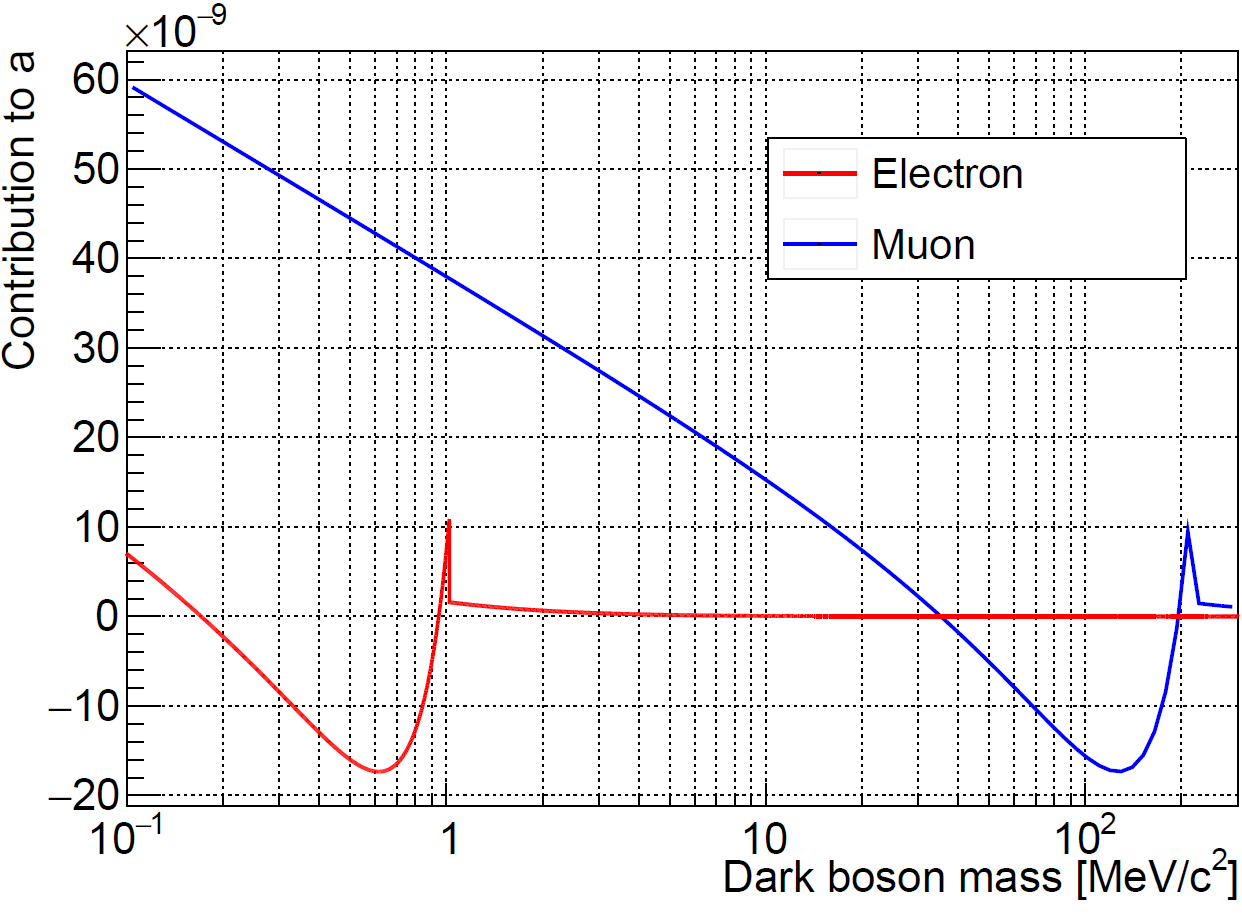}
\includegraphics[width=0.45\textwidth]{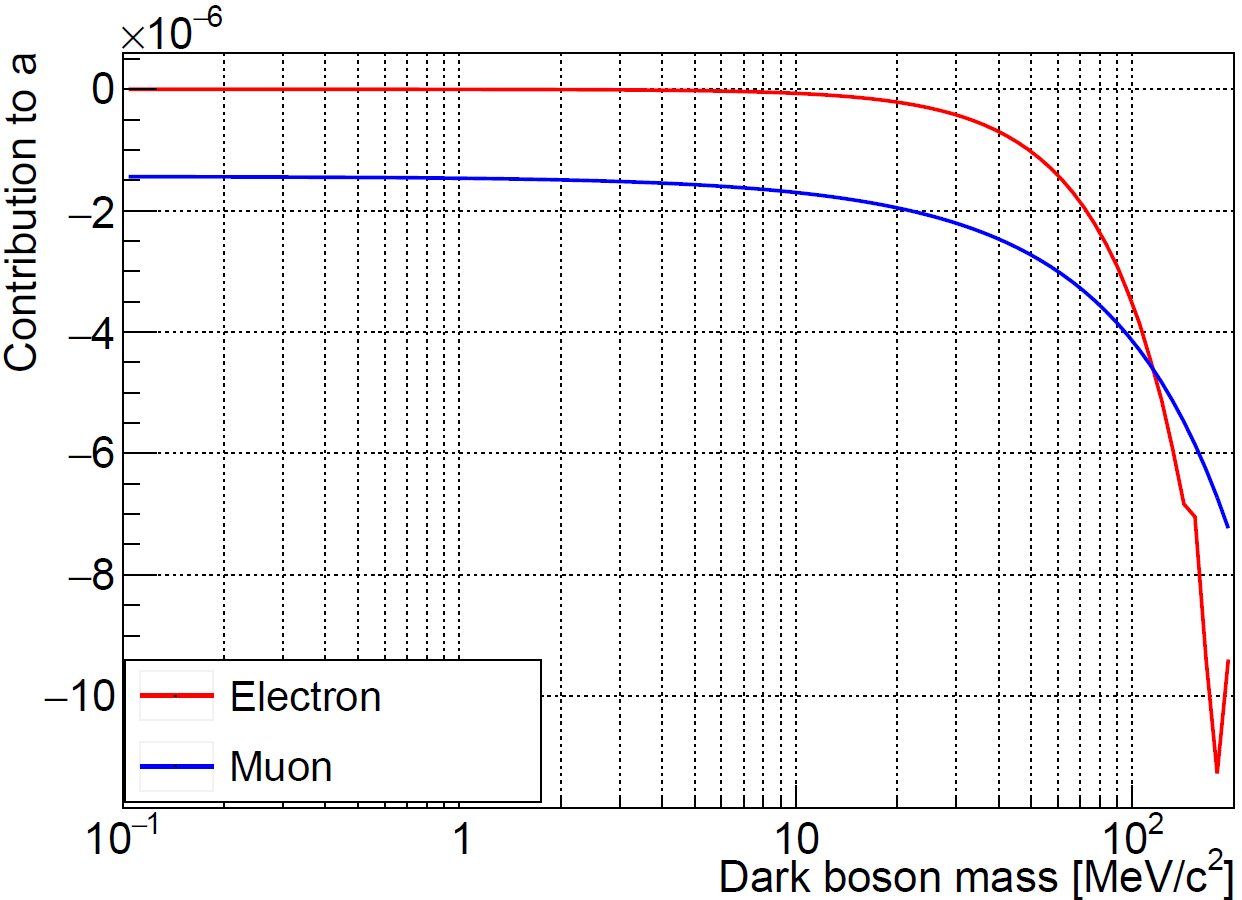}
\centering
\caption{Dependence of the contribution of the dark boson to $a = (g-2)/2$ on the mass of the tensor interacting  dark boson for 
tensor (left) and pseudotensor (right) interaction, $\epsilon_{4/5} = 10^{-3} $.  }
\label{fig:a-mass-dependence}
\end{figure}

The dependence of the $\Delta a_\mu$ correction as a function of the dark boson mass
is shown for the muon and for the electron in fig. \ref{fig:a-mass-dependence} both for
tensor (left) and pseudotensor (right) interactions. 
The contribution as a function of M arising from pseudotensor interaction is always negative as can be seen in fig. \ref{fig:a-mass-dependence} right, 
while for pure tensor interactions there exist
 regions with positive or negative contributions. 
In fact, for M $\leq$ 35 MeV the contrubution to $a$ 
is always positive, while it can be vanishing for 
 $a_{e}$.
%Discontinuities appear at $M = 2 m$ and 
For $M > 2 m$ the contribution of the tensor term is always positive 
and decreases with M while the pseudotensor term leads to a 
negative contribution to the anomalous magnetic moment. 

%Before we continue further, 
%If we consider two dark bosons of mass 20 MeV and 
%equal value of 
%relative coupling strength
%$\epsilon_4 
%= \epsilon_5 
%= 10^{-3}$, the values for $\Delta a$ corrections are:

%\begin{equation}
%    \Delta a_e=1.6\cdot10^{-15}\text{,   } \Delta a_\mu=5.6\cdot10^{-12}\text{ and }\Delta a_\tau=2.2\cdot10^{-11}.
%\end{equation}
%and are dominated by the tensor contribution. 
The presented distributions indicate that with an appropriate choice of the parameters, 
the difference between the experimental and theoretical value for  $\Delta a_\mu$ \cite{AMME, AMMM} can be completely explained by one or more bosons  of mass around 20 MeV with tensor interactions and an interaction constant $\epsilon$ of the order of $10^{-3} - 10^{-4}$.

\begin{figure}[ht]
\includegraphics[width=0.45\textwidth]{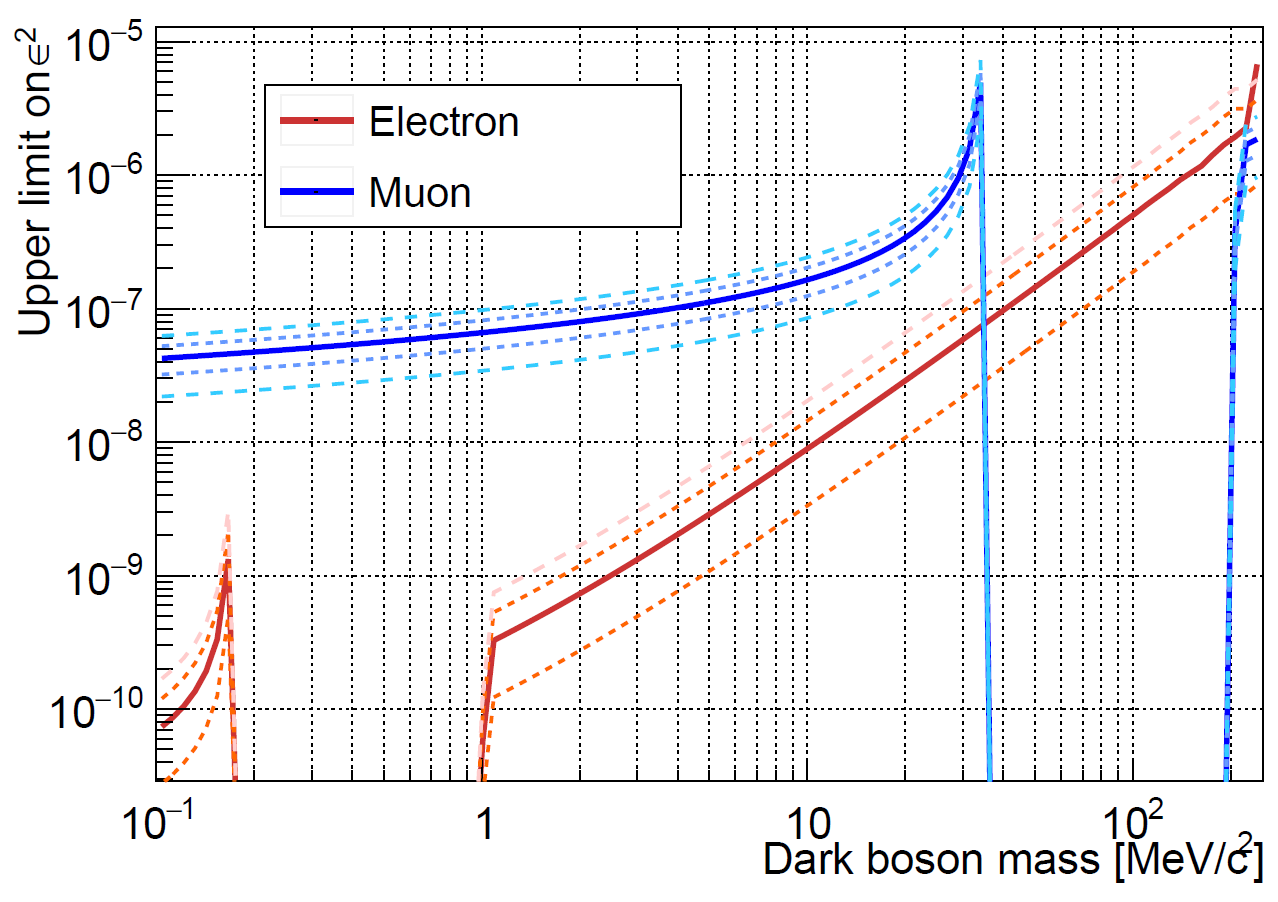}
\centering
\caption{Allowed region of the DB boson parameters assuming  that $A_4$ 
is responsible for the whole contribution to $\Delta a$. 
The solid line denotes the preferred region from the
central value of $\Delta a_{\mu,e}$ while the dashed/dot 
line indicates the $\pm$ 1,2 $\sigma$.}
\label{fig:a4-limits}
\end{figure}

If we assume that the whole discrepancy in $a$ is due to 
a tensor interaction with $A_4$, (i.e. $ \Delta a_{\mu} = \delta_4 \Gamma$), the central values of the 
parameters $\epsilon^2$ and M are given by the 
blue line in fig. \ref{fig:a4-limits}.  
The allowed parameters space covers 
$\epsilon^2$ of the order $\mathcal{O}$($10^{-7} - 10^{-6}$) and M $\leq$ 35 MeV. 
This region is consistent in mass with the 
observed anomaly in $^8 Be$, where $\text{M} \simeq 17 \text{MeV}$, 
while the preferred range of the coupling constant is still to be determined. 
In the presence of additional contribution to 
 $\Delta a_{\mu,e}$ the corresponding lines should be 
 considered as upper limits. However, one should note
 that in the presence of new bosons
 interacting with pseudotensor currents 
 the parameter space is four dimensional and 
 there are regions in which the positive contribution 
 from the tensor interaction is compensated by a negative 
 contribution from the pseudotensor currents.

\section{Sensitivity to $A_1$ and $A_4$ production in positron-on-target annihilation experiments.}

The preferred by the $\Delta a_\mu$ parameter space ($\epsilon^2 \sim \mathcal{O}(10^{-7} - 10^{-6}$) and M$\leq$ 35 MeV) makes it extremely attractive to probe the existence of new light particles 
with (pseudo)tensor interaction with the SM leptons in direct studies of the 
lepton interactions.
Recently, new direction has started in precise study of
the annihilation products of accelerated positrons. 
Such type of experiments are sensitive to $A_{i}$ through the process
$e^+e^-\rightarrow \gamma A_{i}$. 

\begin{figure}[ht]
\includegraphics[width=0.5\textwidth]{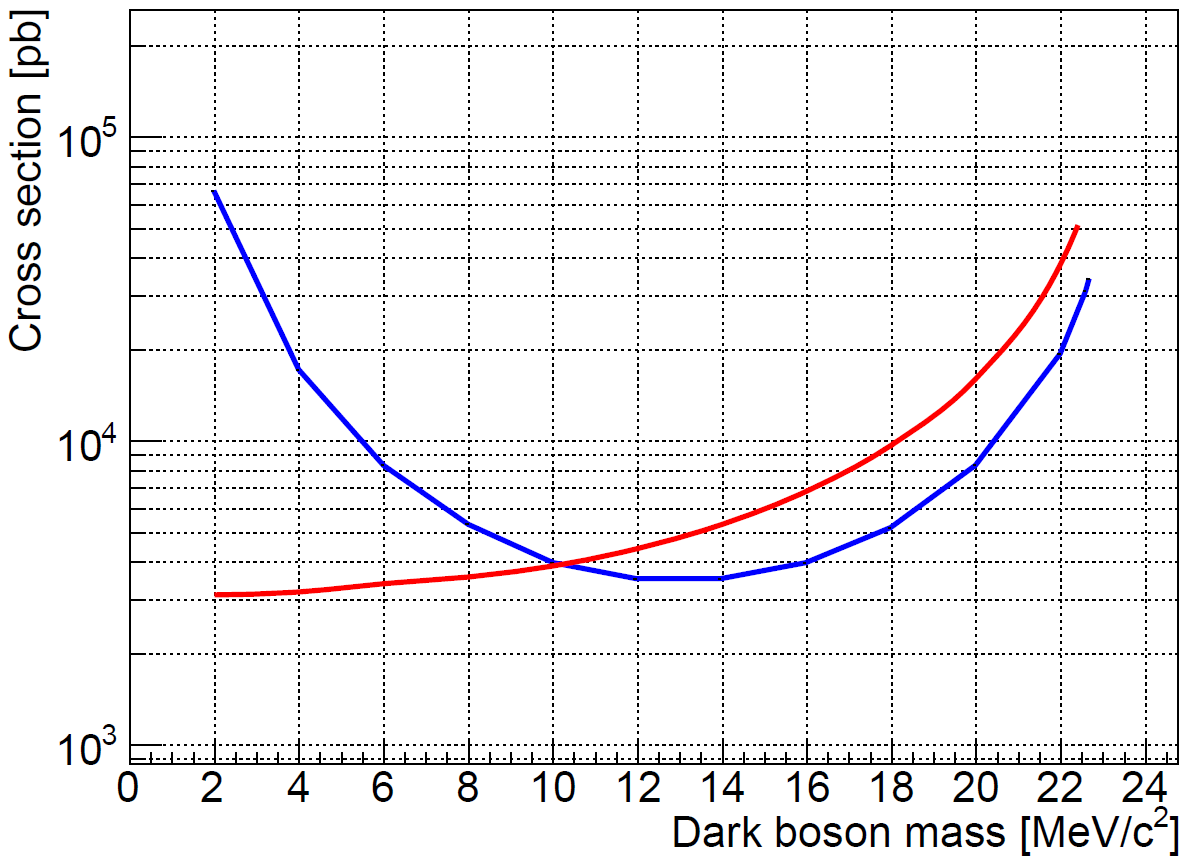}
\centering
\caption{Total cross-sections for the dark boson production in positron-on-target
annihilation, with $p_{e^+} = 550 \text{MeV}$ and $\epsilon = 10^{-3}$ for vector (red) and (pseudo)tensor (blue) interaction.}
\label{fig:cs}
\end{figure}

%From our model we manage to reproduce the result from \cite{PADME}. 
The cross section for the process $e^+e^-\rightarrow \gamma A_1$ follows from the lagrangian (eq. \ref{eq:gen-lagr}) and is

\begin{equation}
    \frac{d\sigma_1}{dz}=\frac{4\pi\epsilon^2\alpha^2}{s}\left(\frac{s-M^2}{2s}\frac{1+z^2}{1-\beta^2z^2}+\frac{2M^2}{s-M^2}\frac{1}{1-\beta^2z^2}\right),
\end{equation}

\begin{equation}
    \sigma_1=\frac{8\pi\alpha^2\epsilon^2}{s}\left[\left(\frac{s-M^2}{2s}+\frac{M^2}{s-M^2}\right)\log\frac{s}{m^2}-\frac{s-M^2}{2s}\right],
\end{equation}

\noindent where $s$ is the invariant mass squared, $\alpha$ is the fine structure constant, $\beta=\sqrt{1-\frac{4m^2}{s}}$ and $z=\cos\theta$.

The differential cross-sections $d\sigma_{4,5}/{dz}$  for the processes $e^+e^-\rightarrow \gamma A_{4,5}$ are identical. 
The values were obtained using CalcHEP. 
%, and defined as
%\begin{equation}
%\begin{gathered}
% \frac{d\sigma_{4,5}}{dz}=\frac{2\pi\epsilon^2\alpha^2}{M^2s^3(1-z^2)^2}\Bigg(8 m^6 z^2+16 m^6 z+8 m^6-8 m^4 M^2 z^2-8 m^4 M^2 z-16 m^4 M^2+4 m^4 s z^3+32 m^4 s z^2+36 m^4 s z+\\
% 8 m^4 s-3 m^2 M^4 z^2 +m^2 M^4-4 m^2 M^2 s z^3-8 m^2 M^2 s z^2-22 m^2 M^2 s z+2 m^2 M^2 s+2 m^2 s^2 z^4+4 m^2 s^2 z^3+\\
% +40 m^2 s^2 z^2+12 m^2 s^2 z+6 m^2 s^2+M^4 s z^3-M^4 s z^2+2 M^4 s z+2 M^4 s-M^2 s^2 z^4-2 M^2 s^2 z^3-\\
% -5 M^2 s^2 z^2-6 M^2 s^2 z-2 M^2 s^2+2 s^3 z^4+12 s^3 z^2+2 s^3\Bigg)
%\end{gathered}
%   \end{equation}
For positron-on-target annihilation with 
positron momentum $p_{e^+} = 550$~MeV, the dependence of the total
cross-section as a function of the dark boson mass is shown in 
fig. \ref{fig:cs}.
The total cross-section $\sigma_{4,5}$
increases both for small masses
M  and when the mass of the dark boson approaches the invariant mass limit ($M\to \sqrt{s}$). 
The behaviour at low M differs significantly for $A_1$ and $A_{4/5}$ 
due to the extra factor $|q|$ in the (pseudo)tensor terms.

\begin{figure}[ht]
\includegraphics[width=0.48\textwidth]{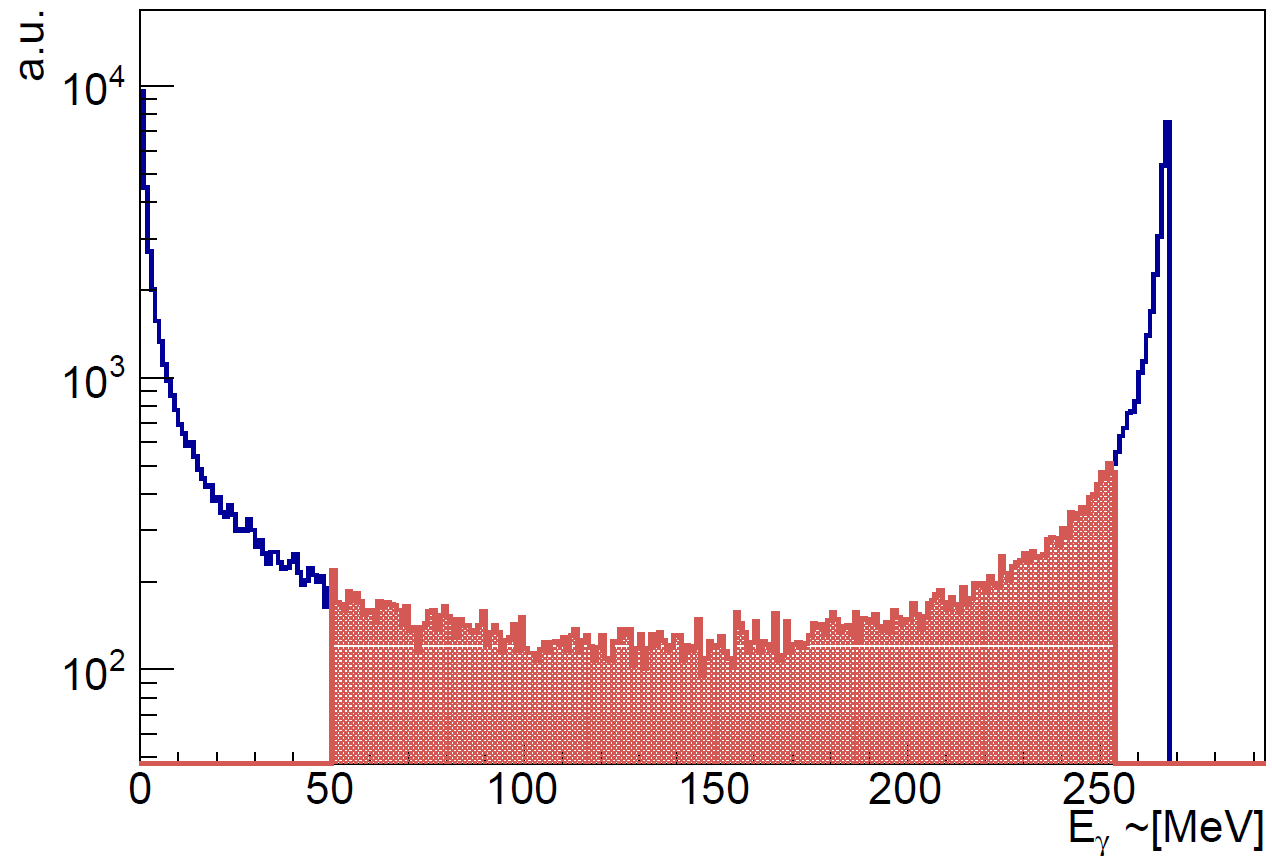}
\includegraphics[width=0.41\textwidth]{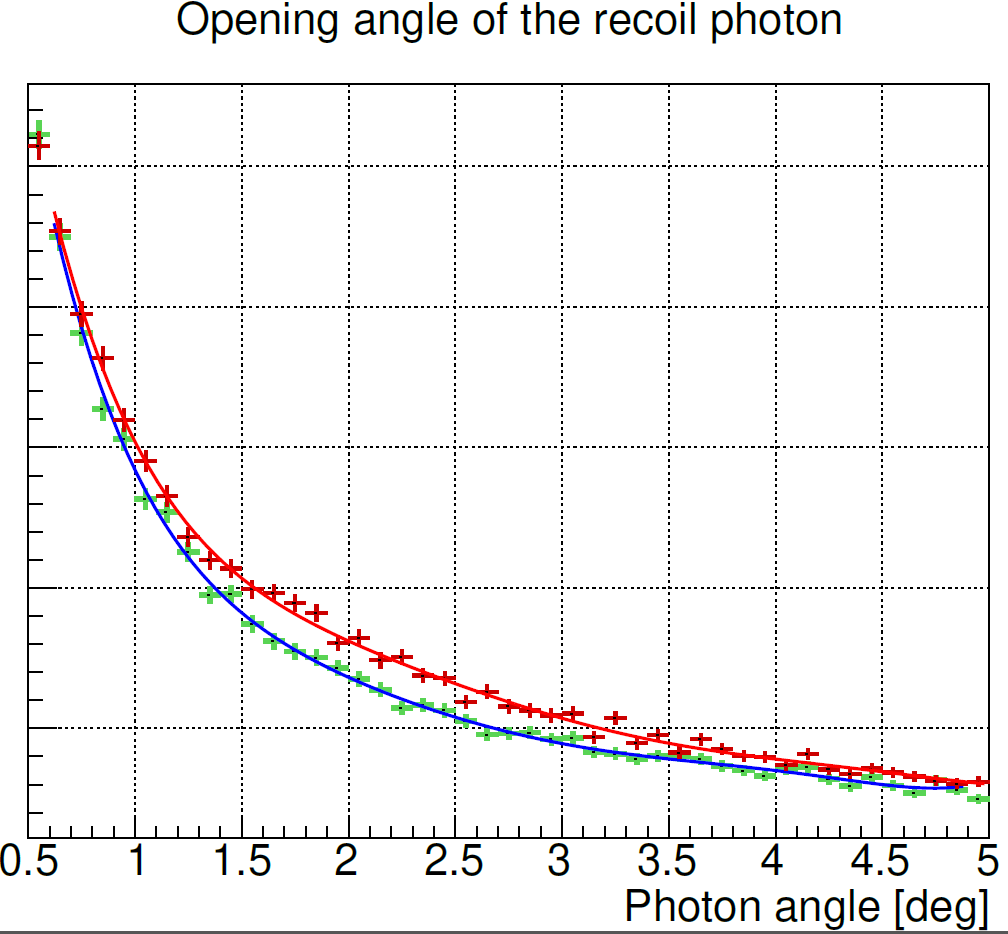}
\centering
\caption{Left: Gamma energy distribution for $e^+e^- \to \gamma + A_4$ with $M$ = 17 MeV for all events (blank histogram) and for the events with the recoil photon in the acceptance of the experimental complex (filled area). Right: Angular distribution of the recoil photon in $e^+e^-\to \gamma A_i$ for vector ($A_1$, green) and tensor ($A_{4/5}$ red) interaction of the dark boson.}
\label{fig:detection}
\end{figure}

The following studies were performed by selecting a 
benchmark point, M = 17 MeV, motivated by the 
existing anomaly in $^8$Be and $^4$He. 
The recoil photon energy distribution is shown in fig. \ref{fig:detection} left. 
Due to its higher mass, most of the initial state energy  
is taken by the dark boson. 
In a typical experiment the recoil photon could only be
detected in a limited opening angle interval \cite{bib:padme}.
The two-body kinematics 
of the events translates the benchmark 
angular interval $ 10~\text{mrad} \leq \theta_{\gamma} \leq 90~ \text{mrad} $ 
to an energy interval $50~\text{MeV} \leq E_{\gamma} \leq 254~\text{MeV}$
of the energy of the photon (shown with filled area). 
Assuming near 100 \% detection efficiency, 
the resulting geometrical acceptance $Acc_{A_{4/5}}$ 
is about 33 \%.

The angular distributions for the vector and (pseudo)tensor case
slightly differ, as can be seen in fig. \ref{fig:detection}, right. 
This difference is mostly pronounced in the region 
$ 20~\text{mrad} \leq \theta_{\gamma} \leq 70~ \text{mrad} $, 
which coincides with the sensitive region 
of the PADME experiment. 
This could allow a single experiment to determine both 
the interaction constants and the type of the interaction 
of the dark boson, in case a positive signal is observed.

%Dependence of the integral in the detector acceptance on the mass of the dark boson. 

\section{Conclusion}
In this work we consider a phenomenological model of interaction between leptons and dark sector particles. 
The study is motivated by the observed 
discrepancy in the anomalous magnetic moment 
of the muon
%anomaly in beryllium-8  decay, 
which interpretation may  be well achieved incorporating a 
new dark sector of particles. 
%There exist experimental evidences that the anomalous beryllium decay might be due to a vector particle, the dark photon, which must have a mass around 17 MeV. 
We investigate the possibility of the existence of a dark boson having a tensor and pseudo-tensor interactions with the fermions. 
%which also can theoretically account for this decay. 
%In a previous work of ours an estimate of the interaction constant was made, which we use in the current work to show that the included dark sector bosons can contribute in a positive way to the understanding of the discrepancy between the theoretical and experimental result for the muon anomalous magnetic moment. 
%Through the nature of the interactions our results incorporate the lepton universality violation which is evident by the fact the the anomalous magnetic moment of the electron is measured to a much greater accuracy compared to the case of a muon.
The proposed interaction 
introduces terms dependent on the lepton mass
beyond just the phase space difference
%violates the Standard model lepton universality 
and may also manifest
itself in the muon anomalous magnetic moment. 
%We hope that our work will broaden the area of particles and interaction types that can be considered as an explanation to a modern problem in high energy physics.
Such a dark boson can be produced in electron-positron interactions,
and be detected in positron-on-target annihilation experiments. 
The described results are applicable to present (e.g. PADME) and 
to future (e.g. SeDS in Brasil \cite{bib:seds} and others \cite{bib:beamdump}) 
positron-on-target annihilation experiments whose sensitivity 
addresses directly the possible simultaneous explanation of 
$\Delta a_{\mu}$ and $^8Be$ anomaly. In addition, a dark boson with mass M $\leq$ 35 MeV can even be probed at hadron and heavy ion 
colliders with detectors allowing access 
to low dilepton invariant mass region, 
for example at ALICE experiment at CERN LHC \cite{bib:alice}. 
While the present work focuses on particular leptonic processes involving (pseudo)tensor 
interactions, numerous different experimental studies 
could potentially be sensitive to the presence of 
$A_{4/5}$ and thus restrict, even significantly limit, the parameter space.

\section*{Acknowledgements}
The authors would like to thank Mihail Chizhov and Tsvetan Vetsov for the valuable discussions. 
This work was partially supported by the Bulgarian National Programme for "Young scientists and Postdocs" and BNSF through COST action CA21106 - COSMIC WISPers in the Dark Universe.

\end{document}